\documentclass[prb,preprint]{revtex4-1}

\usepackage{amsmath}  
\usepackage{amsfonts} 
\usepackage{graphicx} 

\begin{document}

\title{Exploring Gravitational Lensing\footnote{Accepted for publication in \emph{European Journal of Physics} on March 8, 2019. This Accepted Manuscript is available for reuse under a CC BY-NC-ND 3.0 licence after the 12 month embargo period provided that all the terms and conditions of the licence are adhered to.}}

\author{Tilman Sauer}
\email{tsauer@uni-mainz.de} 
\author{Tobias Sch\"utz}
\email{tschuetz@uni-mainz.de}
\affiliation{Institute of Mathematics, Johannes Gutenberg University Mainz, D-55099 Mainz, Germany}

\date{Accepted Version of 19 December 2018.}

\begin{abstract}
In this article, we discuss the idea of gravitational lensing, from a systematic, historical and didactic point of view. We show how the basic lensing equation together with the concepts of geometrical optics opens a space of implications that can be explored along different dimensions. We argue that Einstein explored the idea along different pathways in this space of implication, and that these explorations are documented by different calculational manuscripts. The conceptualization of the idea of gravitational lensing as a space of exploration also shows the feasibility of discussing the idea in the classroom using some of Einstein's manuscripts.
\end{abstract}

\maketitle

\section{Introduction} 

Ever since the double quasar Q0957+561A,B was identified as the first instance of a gravitationally lensed object in 1979, the field has become a highly active and productive field of research with implications for astrophysics, cosmology, exoplanet research, and relativity theory in general.\cite{WambsgansJ1998Lensing} The basic idea of gravitational lensing, however, has a long and interesting prehistory\cite{SchneiderPEtal1992,TrimbleV2001Lenses,RennJEtal2003Eclipses,VallsGabaudD2006Origins} which goes back to early manuscript notes by Einstein dating from the year 1912.\cite{RennJEtal1997Origin} The idea of strong, stellar gravitational lensing arises conceptually from the combination of a law of gravitational light bending interpreted in the framework of geometrical optics as a lens. In general relativity, the law of gravitational light deflection states that the angle of deflection $\alpha$ of a light ray passing a massive astrophysical object (like the sun) of mass $M$ at a distance of closest approach $\rho$ is given by
\begin{equation}
\alpha = \frac{\lambda G M}{c^2\rho},
\label{eq:deflection}
\end{equation}
where $G$ is the gravitational constant and $c$ denotes the velocity of light. In general relativity, the numerical factor $\lambda$ is shown to be $\lambda_{\rm rel}=4$, but Einstein derived the law of light deflection already in 1911 as a consequence of the equivalence hypothesis with a numerical factor of $\lambda_{\rm class}=2$.\cite{EinsteinA1911Einfluss}

Eq.~(\ref{eq:deflection}) can be derived in many ways, both in the relativistic and in the classical Newtonian context, but we will not here be concerned with its derivation. Instead we will take eq.~(\ref{eq:deflection}) as given and explore its consequences. We ask about the optical properties of a gravitational field that is characterized by the law of light deflection (\ref{eq:deflection}) and ask what kind of lensing properties such gravitational field might display. This is what Einstein first did in his notes of 1912, and then again at later occasions. In this paper, we will study this problem in its various possibilities of exploring the implicit features of gravitational lensing.

\section{Derivation of the Lensing Equation}
\begin{figure}[h!]
\centering
\includegraphics[scale=1.0]{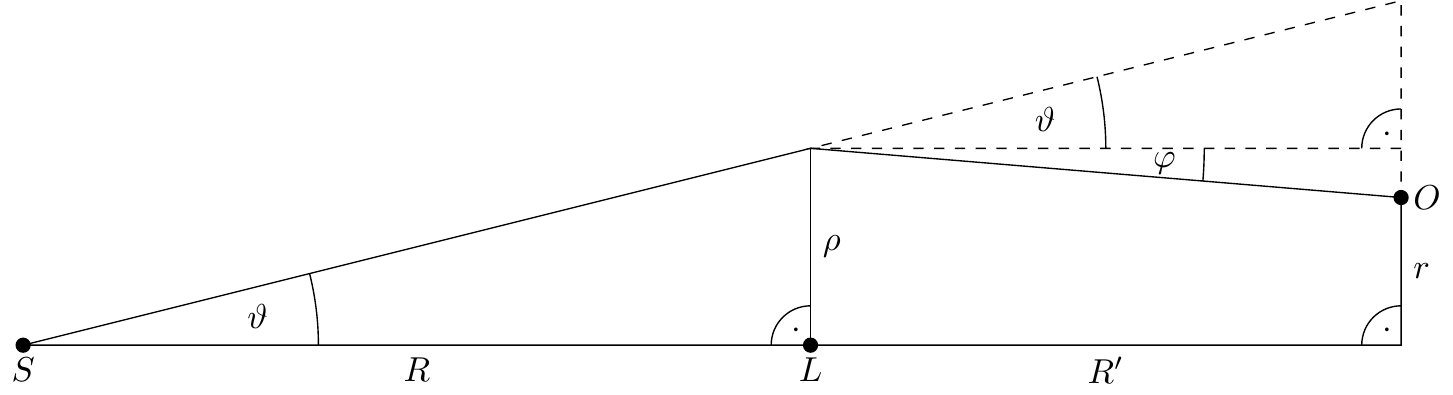}
\caption{Derivation of lensing equation for a light ray emitted by a source $S$, deflected by the gravitational field of a lensing object $L$ and observed by an observer $O$.}
\label{fig:01}
\end{figure}

We want to derive a quadratic equation for the angle under which an observer can see the light emitted by a distant light source behind a massive object that acts as a lens, as displayed in Fig.~\ref{fig:01}. We are given the positions of a light source $S$, a massive object $L$ acting as a lens, and an observer $O$. All three objects are idealized as point-like objects and they uniquely define a plane in Euclidean space. We ask for the condition that an observer can receive light emitted from the source. We idealize the light bending effect of the lens $L$ by assuming that the light ray is bent at one single point according to the law (\ref{eq:deflection}). This idealization corresponds to the thin-lens approximation in geometrical optics. It requires that we define an optical axis which we take to be the straight line connecting light source and lens. The point of deflection is then located on a line orthogonal to the optical axis and passing through the lens. The distance $\rho$ between the point of deflection and the lens determines the angle of deflection according to the law (\ref{eq:deflection}). Extending the optical axis beyond the lens allows us to define also an orthogonal distance $r$ of the observer $O$ from the optical axis. If $R$ denotes the distance between $S$ and $L$, $R'$ denotes the distance between $L$ and the projection of $O$ onto the optical axis. We now ask for a relation between $\rho$ and $r$, i.e. for the condition that a light ray emitted from the source $S$ has to satisfy to reach the eye of an observer $O$. In order to derive this condition, we split the total angle of deflection $\alpha=\vartheta+\varphi$ into two parts such that the common leg of the two angles is parallel to the optical axis. To proceed, we introduce the approximation of small angles $\alpha\approx\sin\alpha\approx\tan\alpha$. We can then write
\begin{equation}
\alpha =  \frac{\lambda G M}{c^2\rho} \equiv \frac{\alpha_0\rho_0}{\rho}= \frac{\rho}{R} + \frac{\rho-r}{R'}.
\label{eq:rho1}
\end{equation}
Here $\alpha_0$ denotes the angle of deflection for some specific, but as yet unspecified distance $\rho_0$, which can be the radius of the sun, the Schwarzschild radius, or the radius of the Einstein ring.
Eq.~(\ref{eq:rho1}) is a quadratic equation for $\rho$ which can also be written as
\begin{equation}
\rho^2 - \frac{R}{R+R'}r\rho - \frac{RR'}{R+R'}\alpha_0\rho_0 = 0.
\label{eq:rho2}
\end{equation}
The two solutions of the quadratic lensing equation correspond to a double image that is seen by an observer. It degenerates into an Einstein ring for the case of perfect alignment of source, lens, and observer. One can also take one further step and derive an expression for the optical magnification.

Note that our derivation of (\ref{eq:rho1}) or (\ref{eq:rho2}) followed the old Cartesian method of analytic geometry: we assumed the solution to a geometric problem solved, as in Fig.~(\ref{fig:01}), then gave names to the relevant lines and angles in the geometric construction and then derived from the figure and problem at hand an algebraic expression that captures the characteristics of the specific problem.\cite{DescartesR1637Geometry} The fact that the quadratic equation, in general, has a second real solution points to a surplus structure of the algebraic representation. Physically, it means that a light ray can reach the observer's eye passing on either side of the lens, although the second possible light path was not part of the original graphical representation of the problem.

Historical research has established that after Einstein's initial investigations of 1912 the basic idea of gravitational lensing and its immediate consequences of double image and Einstein ring as well as geometric amplification were forgotten and rediscovered time and again. A second set of notes by Einstein in the same notebook was shown to date from 1915.\cite{SauerT2008Nova} Some twenty years later, Einstein rederived the idea at the instigation of an amateur scientist, Rudi W. Mandl.\cite{RennJEtal1997Origin,RennJEtal2003Eclipses} A manuscript page with pertinent calculations exists alongside the corresponding publication in a short note in \emph{Science}.\cite{RennJEtal1997Origin,EinsteinA1936Action} Recently, we have also analyzed four sheets of related calculations which are part of a batch of largely unidentified notes and calculations by Einstein,\cite{SchuetzT2018Gravitationslinseneffekt} see Table~\ref{tab:mss}.

\begin{table}[h!]
\centering
\begin{ruledtabular}
\caption{Einstein manuscripts on gravitational lensing. AEA means Albert Einstein Archives at The Hebrew University of Jerusalem. The dating of manuscripts Mss 1--3 to the year 1936 is based on the assumption that they, too, were written around the time when Einstein prepared his publication Ref.~\onlinecite{EinsteinA1936Action}.}
\begin{tabular}{l  l c p{5cm}}
Name & Archival Nr. & Dated & Published \\
\hline	
Prague Notebook 1 &  AEA 3-013 [pp.~43--48] & 1912 & Ref.~\onlinecite[pp.~585--587]{CPAE03} \\
Prague Notebook 2 &  AEA 3-013 [pp.~51--52] & 1915 & Ref.~\onlinecite[p.~589]{CPAE03}  \\
Letter to Mandl & AEA 3-011 [p.~55] & 1936 & Ref.~\onlinecite{RennJEtal1997Origin} \\
German ms for publication & AEA 1-131 & 1936 & www.alberteinstein.info \\
\emph{Science} publication &   & 1936 & Ref.~\onlinecite{EinsteinA1936Action} \\
Ms 1 &  AEA 62-225/368-2 & 1936 & www.alberteinstein.info \\
Ms 2 &  AEA 62-275 & 1936 & www.alberteinstein.info, this article Fig.~\ref{fig:62275}\\
Ms 3 &  AEA 62-349 & 1936 & www.alberteinstein.info \\
\end{tabular}
\label{tab:mss}
\end{ruledtabular}
\end{table}

In addition the idea comes up time and again in a number of publications by other authors over the years.\cite{TrimbleV2001Lenses,RennJEtal2003Eclipses,VallsGabaudD2006Origins}

\begin{figure}[h!]
\centering
\includegraphics[scale=1.0]{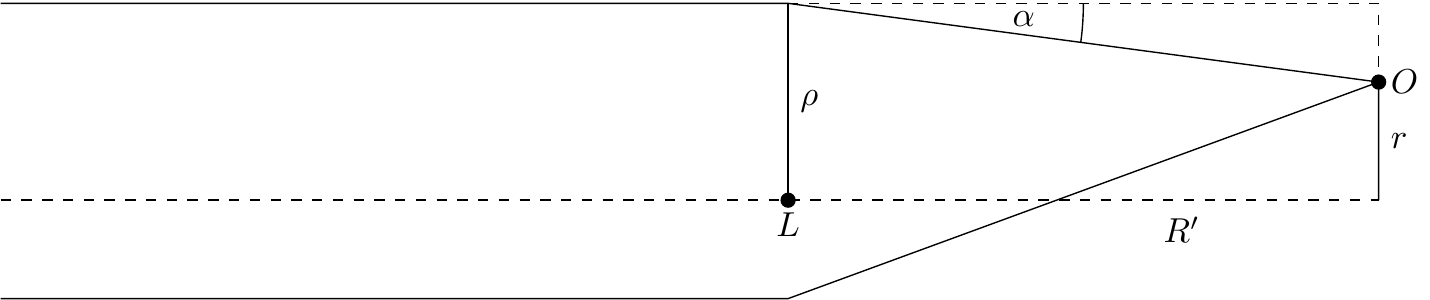}
\caption{Derivation of lensing equation as in Fig.~\ref{fig:01} but with the far source approximation $R\longrightarrow\infty$.}
\label{fig:02}
\end{figure}

In order to analyze and interpret these different manuscript calculations, we have found it useful to think of the problem of (strong) gravitational lensing as a space of conceptual exploration spanned by the law of light deflection and the idea of lensing in geometrical optics. Such a space has different dimensions. From a sketch of the geometric lensing constellation,  one can read off the basic lensing equation, as we have done in our derivation of Eq.~(\ref{eq:rho1}). Going from here one can explore the implications of the lensing idea along different lines: We can normalize the lengths involved so as to obtain a simple form for our quadratic equation (Dimension 1: normalization). Thus, setting
\begin{align}
\tilde{r} &= \sqrt{\frac{R}{(R+R')\alpha_0\rho_0R'}}r \label{eq:norm1},\\
\tilde{\rho} &= \sqrt{\frac{R+R'}{\alpha_0\rho_0R'R}}\rho, \label{eq:norm2}
\end{align}
Eq.~(\ref{eq:rho2}) turns into
\begin{equation}
\tilde{\rho}^2-\tilde{r}\tilde{\rho}-1 = 0.
\label{eq:rho_norm}
\end{equation}
We can also try to simplify the problem by taking the distance between light source and lens (or between lens and observer) to be infinitely long (Dimension 2: far source approximation). On this approximation, Fig.~(\ref{fig:01}) turns into Fig.~(\ref{fig:02}), and the equation we read off from that figure,
along similar lines as above, is
\begin{equation}
\rho^2 - r\rho - \alpha_0\rho_0R' = 0.
\label{eq:rho4}
\end{equation}
Clearly, the same equation is obtained on the algebraic level by taking the limit $R\gg R'$ of Eq.~(\ref{eq:rho2}). Note also that if one takes the $R\gg R'$-limit of (\ref{eq:norm1}) and (\ref{eq:norm2}) and applies it to the lensing equation (\ref{eq:rho4}), one obtains again the normalized form of Eq.~(\ref{eq:rho_norm}).

The following question arises: why do we obtain the same Eq.~(\ref{eq:rho_norm}) if we derive it either by normalizing (\ref{eq:rho2}) directly or by first doing the far source approximation and only then applying the normalization condition?

We can shed light on this issue by asking what would be the graphical representation corresponding to the normalized Eq.~(\ref{eq:rho_norm})? This is the complementary question in the Cartesian program of analytical geometry to the question we asked before. Now, we ask for a geometric construction which realizes an algebraic solution to the problem of gravitational lensing as expressed by Eq.~(\ref{eq:rho_norm}). The solution is found in the Euclidean-Cartesian construction of a quadratic equation.\cite{DescartesR1637Geometry}

\begin{figure}[h!]
\centering
\includegraphics[scale=1.1]{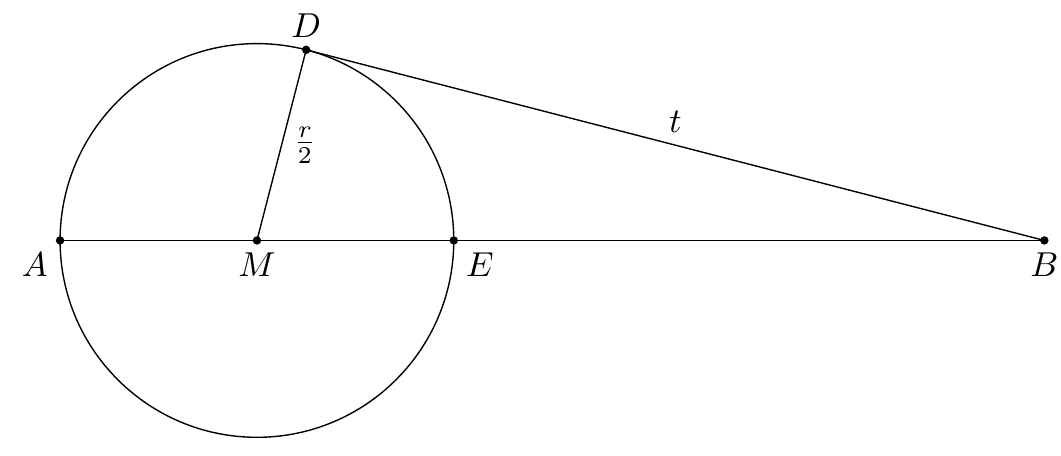}
\caption{Graphical construction of the solutions of Eq.~(\ref{eq:quad}). For $r,t>0$, if $t$ is tangent to the circle $AED$, then $AB$ and $BE=-EB$ are solutions to the quadratic equation.}
\label{fig:quad}
\end{figure}

Consider the quadratic equation
\begin{equation}
\rho^2 - r\rho - t^2 = 0,
\label{eq:quad}
\end{equation}
which corresponds to the geometric construction of Fig.~(\ref{fig:quad}), where the segments $AB$ and $BE=-EB$ represent the two solutions of the quadratic Eq.~(\ref{eq:quad}).\cite{Euclid}
Using this geometric solution to the quadratic equation~(\ref{eq:quad}) allows us to construct a solution to the lensing problem in the following sense. Given Eq.~(\ref{eq:quad}) and distances $R$ and $R'$, we can construct the point of an observer in the imaging plane who would see a double image of the source, consistent with the law of light deflection (\ref{eq:deflection}).
\begin{figure}[h!]
\centering
\includegraphics[scale=1.5]{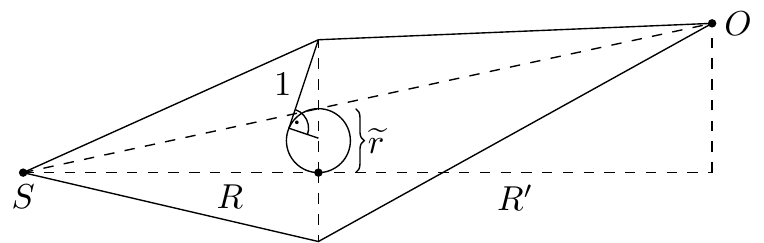}
\caption{Geometric construction of gravitational lensing starting from Eq.~(\ref{eq:rho_norm}). The quadratic equation is geometrically realized in the lensing plane and then projected to the observer's plane.}
\label{fig:lens_constr}
\end{figure}
Fig.~(\ref{fig:lens_constr}) represents the geometric realization of the gravitational lens expressed by the algebraic equation (\ref{eq:rho_norm}), which follows from (\ref{eq:quad}) for $t=1$. In order to do so, we use the geometric construction of Fig.~\ref{fig:quad} to construct the two solutions of the lensing equations in the lensing plane. The smaller solution is added to the lensing plane at the other side of the lens. The location of the observer located on a line orthogonal to the axis, meeting the axis at a distance $R'$ is found by projecting the circle's diameter onto the observer's plane. It can be shown (Appendix \ref{app:1}) that for the observer at $O$, the light deflection indeed follows a law of the form (\ref{eq:deflection}).

Note that here the geometric construction is derived from the algebraic representation and not the other way around as in the derivation of Eq.~(\ref{eq:rho2}). This again corresponds to the Cartesian method where a geometric construction of an algebraic solution was always needed as well.

\begin{figure}[h!]
\centering
\includegraphics[scale=1.3]{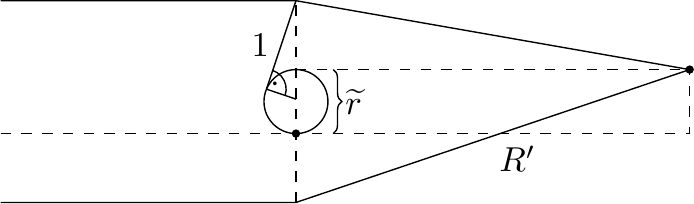}
\caption{Geometric construction of gravitational lensing starting from Eq.~(\ref{eq:rho_norm}) but with the far source approximation $R\longrightarrow\infty$.}
\label{fig:lens_constr2}
\end{figure}

We can now answer the question posed above. For the far source situation we have $R\longrightarrow \infty$ and the incoming light rays in our construction turn into parallels. Then  we can still derive a geometric representation of the quadratic equation, as in Fig.~(\ref{fig:lens_constr2}).

Fig.~(\ref{fig:scheme}) summarizes the foregoing discussion: it shows a square with four corners, which correspond to four different realizations of the lensing equation. The corners are linked by normalization or approximation.
The two dimensions of normalization and far source approximation are indeed completely orthogonal to each other:
\begin{figure}[h!]
\centering
\hspace{-1cm}\includegraphics[scale=0.7]{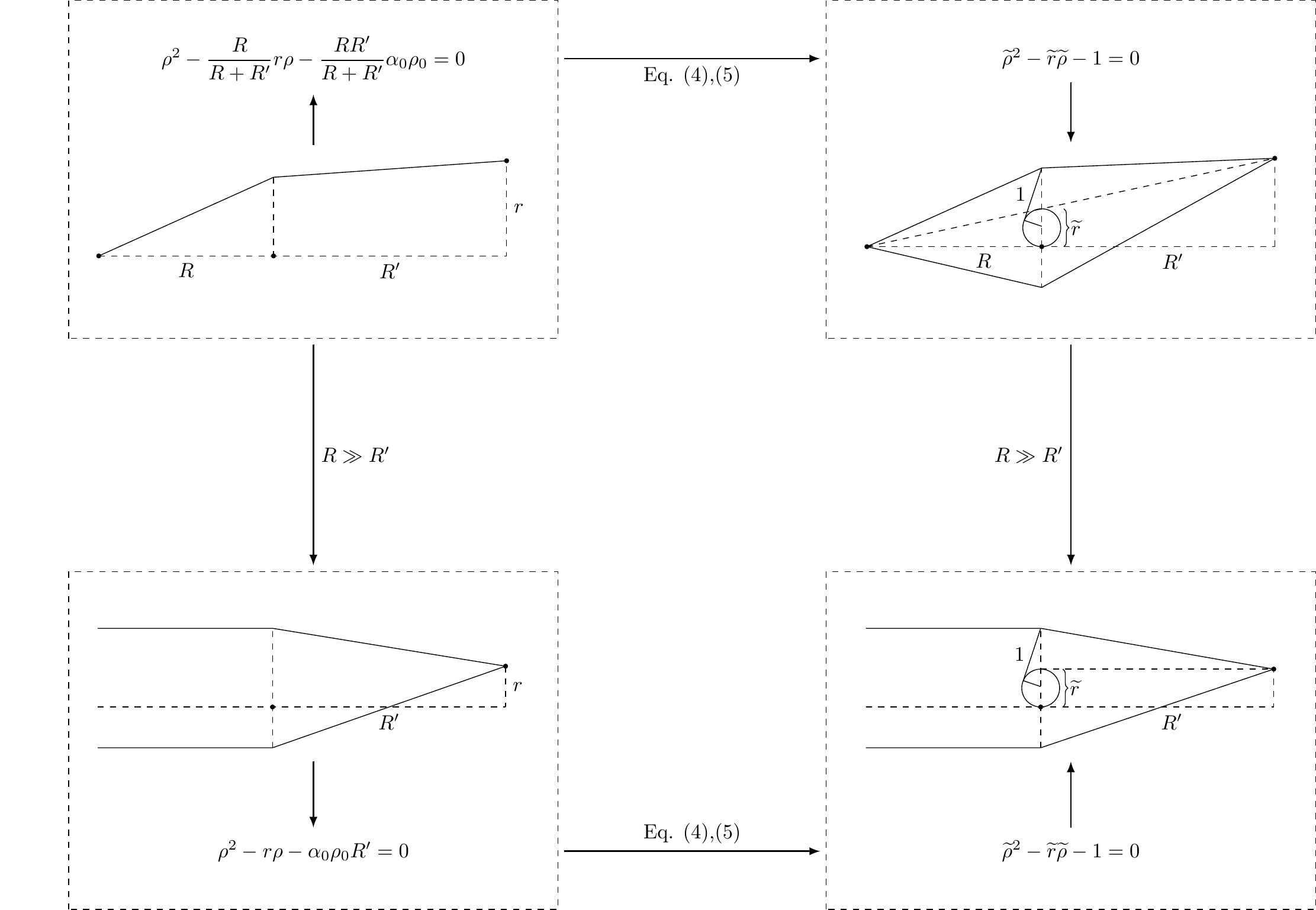}
\caption{The basic idea of gravitational lensing can be represented and explored in different ways. We can either derive a lensing equation from a graphical representation, as displayed on the left hand panels, or we can start from a normalized quadratic lensing equation and represent it geometrically, as shown of the right hand panels. In either case, we can assume finites distances between source, lens, and observer (upper panels) or assume a far-source approximation (lower panels).}
\label{fig:scheme}
\end{figure}
we can either start from a fully dimensional geometric constellation and go to normalized length quantities and then look at the approximate case of one infinite distance to the lens, or we can first do the approximation and then go to normalized variables. 

Starting from any of those four possibilities, we can now proceed along a third dimension and look at the actual solution of the respective quadratic lensing equation to obtain the expression for the double image or the Einstein ring (Dimension 3: solution). For the lensing equation (\ref{eq:rho2}), we obtain the solution
\begin{equation}
\rho_{1,2} = \frac{R}{2(R+R')}r \pm
\sqrt{\left(\frac{R}{2(R+R')}r\right)^2 + \frac{R}{R+R'}\alpha_0\rho_0R'},
\label{eq:quad_sol_full}
\end{equation}
which upon taking the normalization (\ref{eq:norm1}), (\ref{eq:norm2}) simplifies to
\begin{equation}
\tilde{\rho}_{1,2} = \frac{\tilde{r}}{2}\pm\sqrt{\left(\frac{\tilde{r}}{2}\right)^2+1},
\label{eq:quad_sol_norm}
\end{equation}
or which upon taking the far source approximation $R\gg R'$ turns into
\begin{equation}
\rho_{1,2} = \frac{r}{2}\pm\sqrt{\left(\frac{r}{2}\right)^2+\alpha_0\rho_0R'}.
\label{eq:quad_sol_approx}
\end{equation}

Our square of exploration thus turns into a cube. And if we have an explicit quadratic solution, we can also proceed to derive an expression for the magnification. For the full solution (\ref{eq:quad_sol_full}) the amplification $A$ is found to be\cite{SchuetzT2018Gravitationslinseneffekt,HartleJ2003Gravity}
\begin{equation}
A = \frac{\sqrt{\alpha_0\rho_0R'\left(\frac{R+R'}{R}\right)}}{r}
\frac{1+\frac{r^2}{2\alpha_0\rho_0R'}\frac{R}{R+R'}}{\sqrt{1+\frac{r^2}{4\alpha_0\rho_0R'}\frac{R}{R+R'}}}.
\label{eq:amp_full}
\end{equation}
Again, we can apply our normalization conditions (\ref{eq:norm1}), (\ref{eq:norm2}) to obtain
\begin{equation}
A = \frac{1+\frac{\tilde{r}^2}{2}}{\tilde{r}\sqrt{1+\frac{\tilde{r}^2}{4}}},
\label{eq:amp_norm}
\end{equation}
or, alternatively, we take the far source approximation $R\gg R'$ and obtain
\begin{equation}
A = \frac{\sqrt{\alpha_0\rho_0R'}}{r}
\frac{1+\frac{r^2}{2\alpha_0\rho_0R'}}{\sqrt{1+\frac{r^2}{4\alpha_0\rho_0R'}}}.
\label{eq:amp_approx}
\end{equation}

\begin{figure}[h!]
\centering
\includegraphics[scale=0.9]{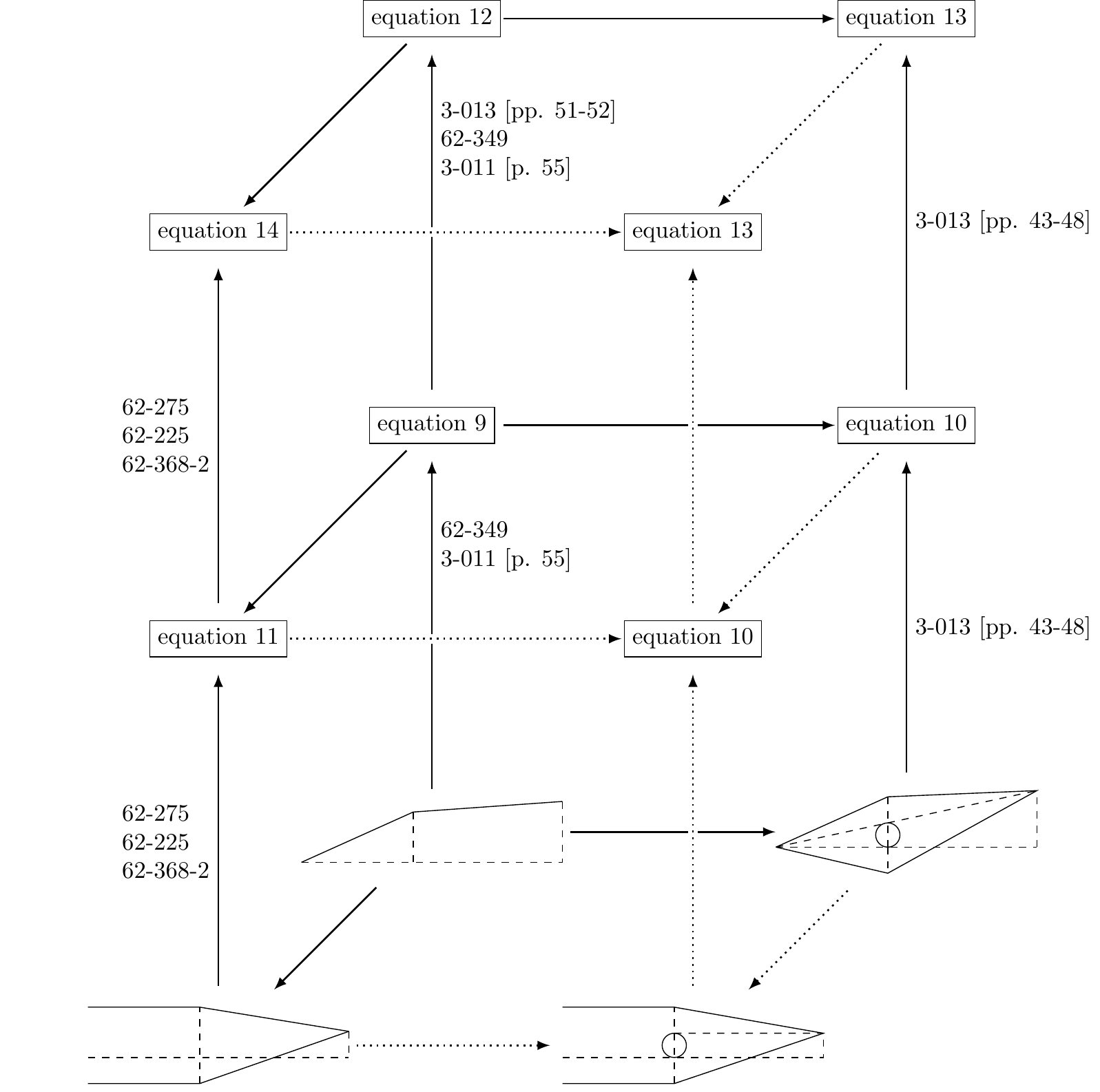}
\caption{Exploration space for probing derivations and implications of the lensing equation along different lines. Starting from one of the four possible representations of the lensing equation (Fig.~\ref{fig:scheme}), which here form the bottom layer, one can solve the equations to find the angle of the double image (first layer) and from there obtain the geometric amplification (second layer). As indicated next to the vertical arrows, in his notes (Table~\ref{tab:mss}) Einstein proceeded along different pathways in this space.}
\label{fig:cube}
\end{figure}

With this step we add another layer to our cube. The situation is visualized in Fig.~(\ref{fig:cube}).

\section{Einstein's derivations}

\begin{figure}[h!]
\centering
\includegraphics[scale=0.9]{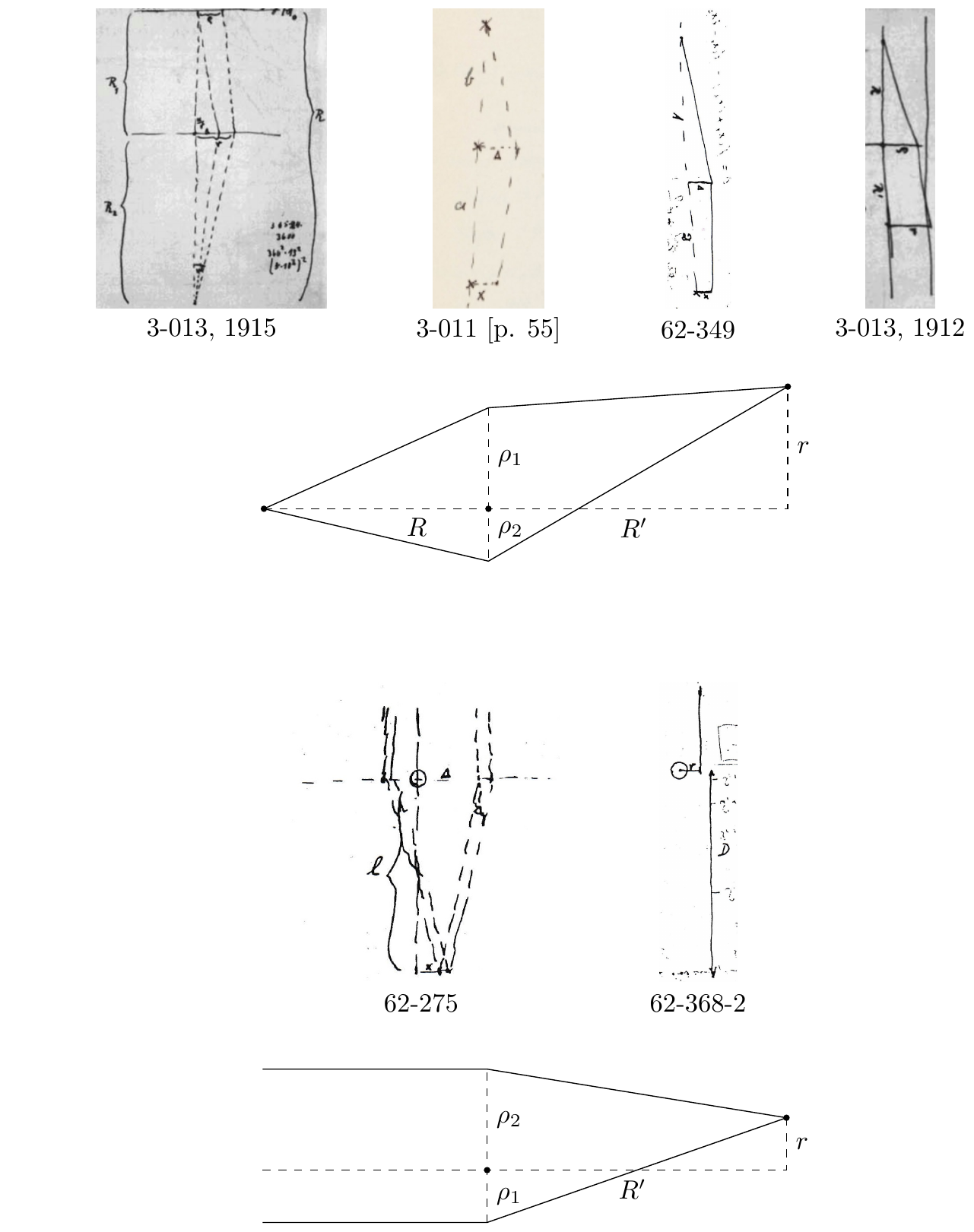}
\caption{As the graphical sketches (\copyright Albert Einstein Archives, The Hebrew University of Jerusalem, Israel), which accompany the various derivations of the lensing equation in Einstein's notes, show, he either started out from the top left corner of Fig.~\ref{fig:scheme} (the far left bottom corner of Fig.~\ref{fig:cube}), or from the far source approximation, i.e.\ the bottom left corner of Fig.~\ref{fig:scheme} (the front left bottom corner of Fig.~\ref{fig:cube}). }
\label{fig:zuordnung}
\end{figure}

The preceding systematic discussion allows us to identify the various calculational notes by Einstein that deal with gravitational lensing as different pathways in the space of exploration as depicted in Fig.~\ref{fig:cube}.
To begin with, we note that Einstein starts either from the far left corner or from the front left corner. This is illustrated by the different sketches accompanying his calculations as shown in Fig.~\ref{fig:zuordnung}.

We can now identify Einstein's various calculations as pathways in our space of exploration of Fig.~\ref{fig:cube}.
For instance, in his earliest documented notes (AEA~3-013, pp.~43--48), Einstein started out from the far left corner, then immediately moved to the far right corner by normalizing, and then moved upwards by computing the quadratic solution and the amplification. When he came back to the problems two years later in the same notebook, he did not do the normalization but rather proceeded directly upwards from the far left corner.

The same route (always with different notation, to be sure) was taken in 1936 in his notes of the ``Letter to Mandl'', as well as in AEA~62-349. Three other sheets from that period, however, proceed by starting directly from the far source approximation and moving upwards in Fig.~\ref{fig:cube} from there. Finally, we observe that in his brief published note,\cite{EinsteinA1936Action} Einstein does not indicate at all his way of proceding. Instead he only gives the final result for the amplification, which corresponds to Eq.~(\ref{eq:amp_approx}). In his note to \emph{Science}, Einstein assumes a far source approximation.
\bigskip
\bigskip

\begin{figure}[h!]
\centering
\includegraphics[scale=0.8]{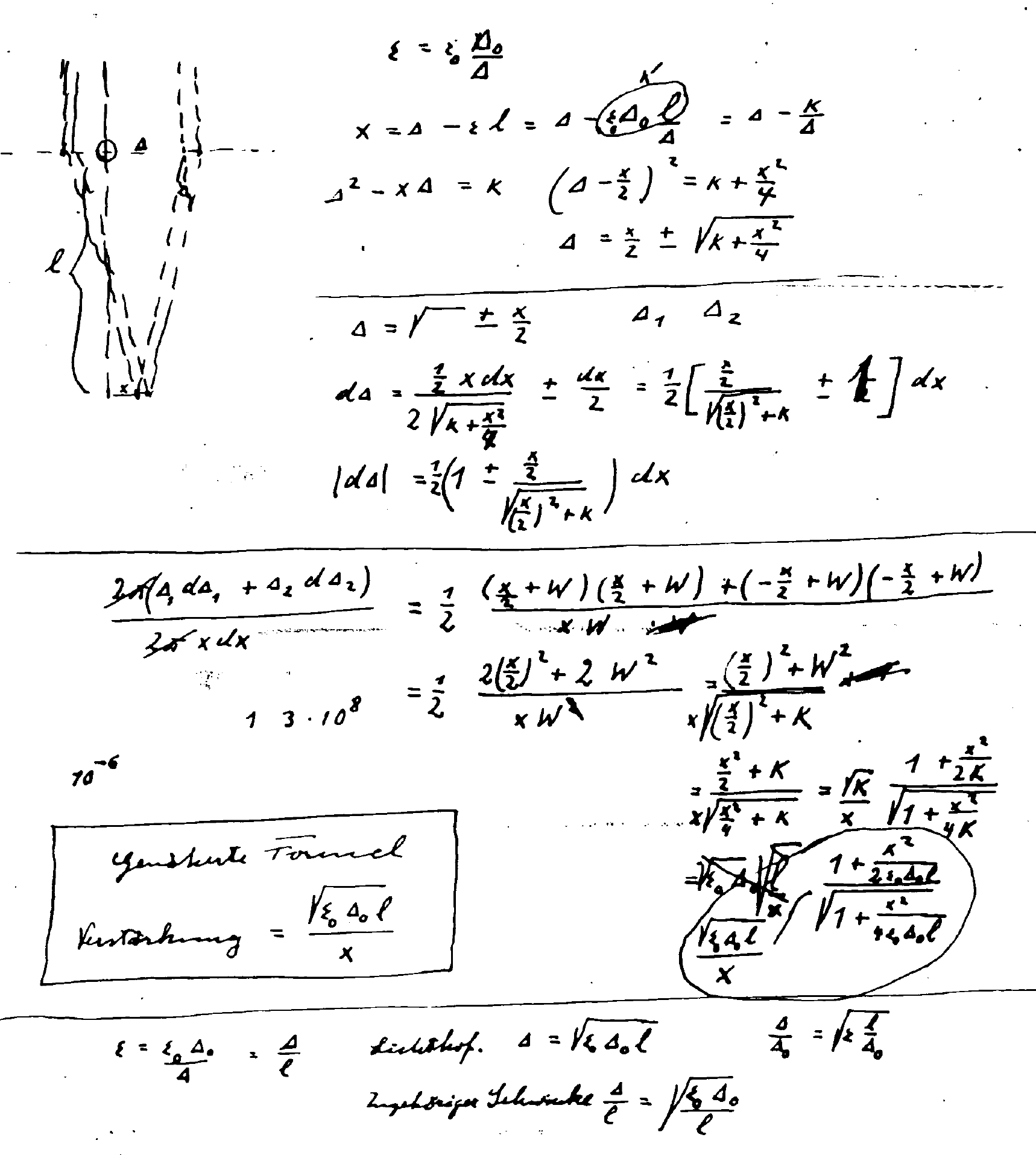}
\caption{Einstein's derivation of the lensing equation, solution, and amplification in AEA 62-275 (\copyright Albert Einstein Archives, The Hebrew University of Jerusalem, Israel). In this calculation, Einstein proceeds by moving upwards from the front left bottom corner of Fig.\ref{fig:cube}.}
\label{fig:62275}
\end{figure}

In Fig.~(\ref{fig:62275}), we present a page with calculations by Einstein that is part of a large batch of manuscripts.
As with all other derivations, the notation is idiosyncratic but conceptually the derivation follows a pathway in Fig.~\ref{fig:cube}.
The sketch in the top left corner clearly indicates that Einstein assumes a far source approximation from the outset with light rays converging to an observer placed at a distance $x$ from the optical axis and a distance $l$ from the lensing mass, i.e.\ to map this instance of his calculations to our general discussion, we set $r\rightarrow x$,  $R'\rightarrow l$, $\rho\rightarrow \Delta$, $\alpha\rightarrow \epsilon$, $\alpha_0\rightarrow \epsilon_0$, $\rho_0\rightarrow\Delta_0$. Einstein reads off the lensing equation 
\begin{equation}
x = \Delta -\epsilon_0\Delta_0\frac{l}{\Delta}
\end{equation}
which he immediately transforms by setting
\begin{equation}
\kappa \equiv \epsilon_0\Delta_0 l
\end{equation}
to obtain the simple form
\begin{equation}
\Delta^2 -x\Delta = \kappa.
\label{eq:quad_eq}
\end{equation}
Quadratic completion immediately provides the solution
\begin{equation}
\Delta = \frac{x}{2} \pm \sqrt{\kappa+\frac{x^2}{4}}.
\label{eq:quad_sol}
\end{equation}
The calculation therefore instantiates a path starting from the front left bottom corner upwards in Fig.~\ref{fig:cube}.
For the derivation of light amplification of that sheet, see Appendix \ref{app:2}.

\section{Exploring gravitational lensing in the classroom}

The exploration of the conceptual space depicted in Fig.~(\ref{fig:cube}) is entirely accessible at the high school level. In a short proof-of-concept project we used Einstein's notes to introduce the basic ideas of gravitational lensing in a physics course in twelfth grade of a local German Gymnasium. In three consecutive lessons, each lasting 45 minutes, the students were introduced to the geometry of a typical gravitational lensing situation. They also set up the lensing equation~(\ref{eq:quad_eq}) and solved it to obtain the solutions in Eq.~(\ref{eq:quad_sol}) by reconstructing and interpreting Einstein's notes. We used the first four rows of Fig.~(\ref{fig:62275}) including the sketch. The project presupposed with the students a familiarity of a) basic notions of geometrical optics, notably the thin-lens approximation, b) basic Euclidean geometry and trigonometry using the small angle approximation and the assumption of parallel incoming light rays, and c) manipulations of a quadratic equation. At the same time, the problem of gravitational lensing provides a historical context that allows students to explore these topics in ways that differ from the standard paradigm.

\begin{figure}[h!]
\centering
\includegraphics[scale=0.5]{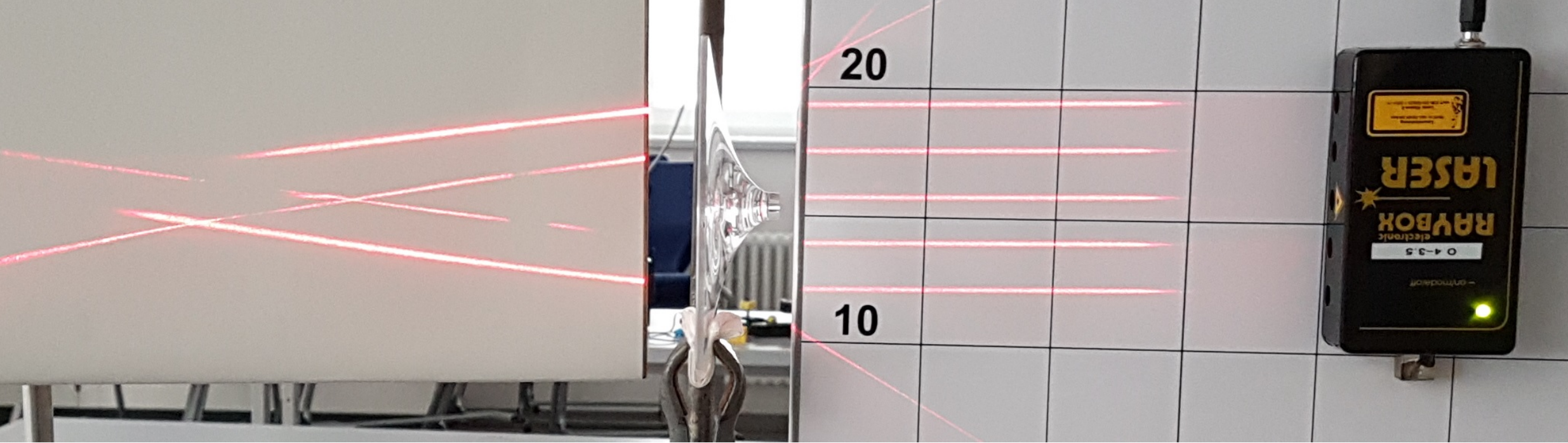}
\caption{Simulation of the geometry of gravitational lensing using an optical lens with the shape of the foot of a wine glass.}
\label{fig:simul}
\end{figure}

In the first lesson students were introduced to the geometry of gravitational lensing using a model gravitational lens made from glass, which looks like the foot of a wine-glass.\cite{RefsdalSEtal1994Lenses,SurdejJEtal2010Telescope} In the second lesson the students used Einstein's notes to deduce the lensing equation from the geometrical situation and computed the solutions whose geometrical interpretation they discussed. We found that the authenticity of Einstein's calculations and the challenge to explore his thinking directly from his notes very much added to the motivation of the students. This has been made evident by the students' answers and comments on evaluation sheets, which they filled in after the project was completed.\cite{SchuetzT2018Gravitationslinseneffekt} The third lesson helped the students to consolidate their understanding by contrasting the model gravitational lens with the geometric modeling, investigating, for example,  the dependency of the derived Einstein radius on the distance between observer and lens by looking at a candle through the model lens. The example, furthermore, led the students to appreciate the lensing idea both in its similarities and dissimilarities to the usual convex or concave lenses, see Fig.~(\ref{fig:simul}).

We propose that if the teaching schedule or circumstances allow for more time, not only can the first part of Einstein's calculations prove suitable for a discussion at high school level, but the remainder of Fig.~(\ref{fig:62275}) concerning the derivation of the magnification factor as well. For this more ambitious project, students need to have some basic knowledge about infinitesimal calculus, need to learn about or be familiar with solid angles as well as with the notion of brightness and geometric light amplification. As a guide, the relevant calculations of geometric light amplification for the case of Fig.~(\ref{fig:62275}) have been spelled out in Appendix~\ref{app:2}.

Furthermore, all the different possible pathways laid out in Fig.~(\ref{fig:cube}) can function as alternative approaches to explore the idea of gravitational lensing, at this level of strong, stellar lensing. And, as indicated, most of the pathways are documented by Einstein's calculations and can thus be explored by students at this level. Especially starting from the far left corner and computing the full lensing equation~(\ref{eq:rho2}) and the solutions in equation~(\ref{eq:quad_sol_full}) is, from a conceptual point of view, easier to comprehend since the far source approximation and therefore the assumption of parallel incoming light rays is not needed. However, the formulas obtained then look more complicated, which could demand more time end efforts from the students.  In summary, various manuscript calculations by Einstein have been published (see Table~\ref{tab:mss}) and allow students to explore and reconstruct Einstein's ideas and the concept of gravitational lensing directly from his own publication and scratch notes. Our proposal thus contributes to a broader program of utilizing history and philosophy of science for science teaching.\cite{MatthewsM2014Handbook}

\section{Concluding remarks}

We have reconstructed and compared various notes by Einstein in which he derived the basic equation of gravitational lensing and computed its solutions as well as the corresponding expression for light amplification. Apart from differences in notation, these various derivations proceed along different pathways in a space of exploration which is spanned by three mutually orthogonal dimensions of normalization, approximation, and solution. On this level, gravitational light bending is a phenomenon in flat Euclidean space but the very same phenomenon also prepares students for a study of curved space-time in a relativistic context.

\appendix

\section{Geometric construction of light deflection law}
\label{app:1}

\begin{figure}[h!]
\centering
\includegraphics[scale=1.4]{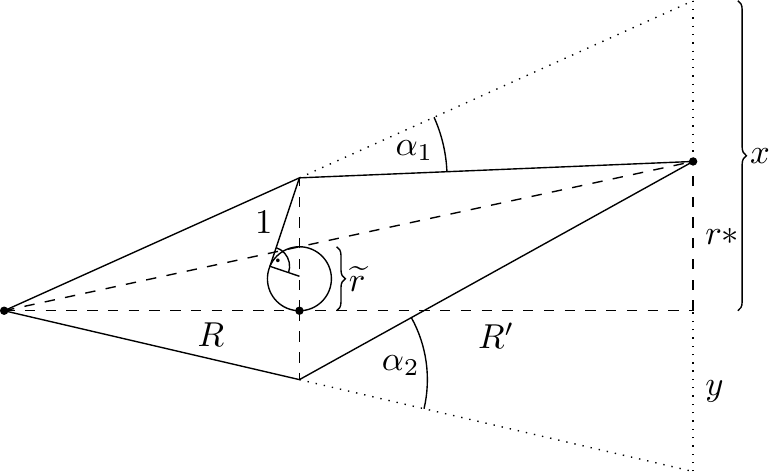}
\caption{Geometric construction of gravitational lensing starting from Eq.~(\ref{eq:rho_norm}). The quadratic equation is geometrically realized in the lensing plane and then projected to the observer's plane.}
\label{fig:lens_constr3}
\end{figure}

We wish to show that the geometric construction of Fig.~\ref{fig:lens_constr} realizes a light bending law proportional to $1/\tilde{\rho}$. We complete Fig.~\ref{fig:lens_constr} as shown in Fig.~\ref{fig:lens_constr3}.
In the small angle approximation, if $\tilde{\rho}_{1,2}$ are the solutions of Eq.~(\ref{eq:rho_norm}), the upper light ray is deflected by an angle
\begin{equation}
\alpha_1\approx \frac{x-r^{\star}}{R'}
= \frac{x-r^{\star}}{R+R'} \frac{R+R'}{R'}
= \frac{\tilde{\rho}_1-\tilde{r}}{R}\frac{R+R'}{R'}
= |\tilde{\rho}_2|\frac{R+R'}{RR'}
= \frac{t^2}{\tilde{\rho}_1}\frac{R+R'}{RR'},
\end{equation}
while the lower light ray is deflected by an angle
\begin{equation}
\alpha_2\approx \frac{y+r^{\star}}{R'}
= \frac{y+r^{\star}}{R+R'} \frac{R+R'}{R'}
= \frac{|\tilde{\rho}_2|+\tilde{r}}{R}\frac{R+R'}{R'}
= \tilde{\rho}_1\frac{R+R'}{RR'}
= \frac{t^2}{|\tilde{\rho}_2|}\frac{R+R'}{RR'}.
\end{equation}
Here we have used $\tilde{\rho}_1-|\tilde{\rho}_2|=\tilde{r}$ and $\tilde{\rho}_1\cdot|\tilde{\rho}_2|=t^2$.
It follows that both light rays are deflected according to the law Eq.~(\ref{eq:rho1}) upon identifying
\begin{equation}
\alpha_0\tilde{\rho}_0 \equiv t^2\frac{R+R'}{RR'}.
\end{equation}

\section{Reconstruction of AEA 62-275}
\label{app:2}

We reconstruct Einstein's derivation of the light amplification in AEA 62-275 (Fig.~\ref{fig:62275}), see Ref.~\onlinecite[pp.~62--66]{SchuetzT2018Gravitationslinseneffekt}, for details.

Einstein denotes the two solutions of Eq.~(\ref{eq:quad_sol}) by $\Delta_1$ and $\Delta_2$. Differentiation with respect to $x$ yields
\begin{equation}
d\Delta = \frac{\frac{1}{2}x\,dx}{2\sqrt{\kappa+\frac{x^2}{4}}} \pm \frac{dx}{2}
= \frac{1}{2}\left[
\frac{\frac{x}{2}}{\sqrt{\left(\frac{x}{2}\right)^2+\kappa}}\pm 1
\right]dx.
\end{equation}
Setting
\begin{equation}
W = \sqrt{\left(\frac{x}{2}\right)^2+\kappa},
\end{equation}
Einstein computes the relative amplification by looking at the ratio of the areas of differential surface elements, i.e.\ cross sections of light ray bundles, both in the lensing plane and in the observer plane. The calculation goes like
\begin{align}
\frac{2\pi(\Delta_1\,d\Delta_1+\Delta_2\,d\Delta_2)}{2\pi x\,dx}
&= \frac{1}{2}\frac{\left(\frac{x}{2}+W\right)\left(\frac{x}{2}+W\right)+\left(-\frac{x}{2}+W\right)\left(-\frac{x}{2}+W\right)}{xW} \\
&= \frac{1}{2}\frac{2\left(\frac{x}{2}\right)^2+2W^2}{xW} = \frac{\left(\frac{x}{2}\right)^2+W^2}{x\sqrt{\left(\frac{x}{2}\right)^2+\kappa}} \\
&= \frac{\frac{x^2}{2}+\kappa}{x\sqrt{\left(\frac{x}{2}\right)^2+\kappa}} = \frac{\sqrt{\kappa}}{x}\frac{1+\frac{x^2}{2\kappa}}{\sqrt{1+\frac{x^2}{4\kappa}}} \\
&= \frac{\sqrt{\epsilon_0\Delta_0l}}{x}\frac{1+\frac{x^2}{2\epsilon_0\Delta_0l}}{\sqrt{1+\frac{x^2}{4\epsilon_0\Delta_0l}}}.
\end{align}
In a box, Einstein wrote ``approximated formula'' (``Gen\"aherte Formel'') and
\begin{equation}
\text{``Amplification'' (``Verst\"arkung'')} = \frac{\sqrt{\epsilon_0\Delta_0l}}{x}.
\end{equation}
This expression is obtained by the approximation
\begin{equation}
1+\frac{x^2}{2\epsilon_0\Delta_0l} \approx \sqrt{1+\frac{x^2}{4\epsilon_0\Delta_0l}}.
\end{equation}
Since $\epsilon_0\Delta_0$ is a constant, the approximation amounts to the assumption $x^2\ll l$.

Finally, Einstein considers the situation
\begin{equation}
\epsilon = \frac{\epsilon_0\Delta_0}{\Delta}=\frac{\Delta}{l}.
\end{equation}
Since the angle of deflection can also be written as $\epsilon=(\Delta-x)/l$, one can now look at the case $x=0$, i.e.\ the case of the Einstein ring. The radius then follows from Eq.~(\ref{eq:quad_sol}) as $\Delta=\sqrt{\epsilon_0\Delta_0l}$, and Einstein comments on this expression by writing the term ``halo'' (``Lichthof'').

\end{document}